\documentclass[preprint,aps,prl]{revtex4-1}

\usepackage{graphicx}	
\usepackage{dcolumn}	
\usepackage{bm}		

\begin{document}
\preprint{JFE}


\title{Magnetothermodynamics: 
measuring the equations of state of a compressible magnetized plasma}
\author{M. R. Brown}
\email{doc@swarthmore.edu}
\author{M. Kaur}
\affiliation{Department of Physics and Astronomy \\
Swarthmore College \\
Swarthmore, PA 19081-1397} 

\date{\today}

\begin{abstract}
Magnetothermodynamics (MTD) is the study of compression and expansion of magnetized plasma with an eye towards identifying equations of state for magneto-inertial fusion experiments. We present recent results from SSX experiments on the thermodynamics of compressed magnetized plasmas. In these experiments, we generate twisted flux ropes of magnetized, relaxed plasma accelerated from one end of a $1.5~m$ long copper flux conserver, and observe their compression in a closed conducting boundary installed at the other end. Plasma parameters are measured during compression. The instances of ion heating during compression are identified by constructing a PV diagram using measured density, temperature, and volume of the magnetized plasma. The theoretically predicted MHD and double adiabatic (CGL) equations of state are compared to experimental measurements to estimate the adiabatic nature of the compressed plasma. Since our magnetized plasmas relax to an equilibrium described by magnetohydrodynamics, one might expect their thermodynamics to be governed by the corresponding equation of state. However, we find that the magnetohydrodynamic equation of state is not supported by our data. Our results are more consistent with the parallel CGL equation of state suggesting that these weakly collisional plasmas have most of their proton energy in the direction parallel to the magnetic field. 
\end{abstract}


\maketitle


\section{\label{sec:Intro}Introduction}

Magnetothermodynamics (MTD) is the study of compression and expansion of magnetized plasma with an eye towards identifying equations of state.  The physics of MTD was first elucidated at the SSX MHD wind tunnel at Swarthmore College \cite{BrownJPP,BrownPSST}.  In experiments using relaxed Taylor states \cite{Taylor74,Taylor86,Cothran09,Gray13}, we accelerated and compressed magnetized plasmas while measuring local proton temperature $T_i$, plasma density $n_e$, and magnetic field ${\bf B}$.  A particular equation of state (EOS) was identified in these experiments \cite{KaurPRE,KaurJPP}. 

Magneto-inertial fusion (MIF) experiments rely on the compression and heating of magnetized plasmas \cite{Lindemuth95,Wurden16}.  The compression is often performed mechanically by either physically imploding a liner \cite{Degnan13}, or collapsing a liquid metal wall \cite{Laberge09}.  As such, the magnetic lifetime of the magnetized plasma state should be much longer than the mechanical implosion time.  Recently, on a smaller scale ($5~mm$ diameter), compression experiments were carried out at the magnetized liner inertial fusion (MagLIF) experiment at Sandia.  In these experiments, plasmas are heated in $100~ns$ from $100~eV$ to $4~keV$, and magnetic fields are amplified from $10~T$ to $1000~T$ \cite{Gomez14,Gomez15}.  Each of these experiments could benefit from a well-established equation of state.

In section II we review the SSX experiment and diagnostics, in section III we discuss the experiment and results, finally in section IV we propose a fusion engine using Taylor state compression.

\section{\label{sec:SSX}SSX plasma wind tunnel and diagnostics}

\begin{figure*}[!h]
\begin{center}
\includegraphics[width=5in]{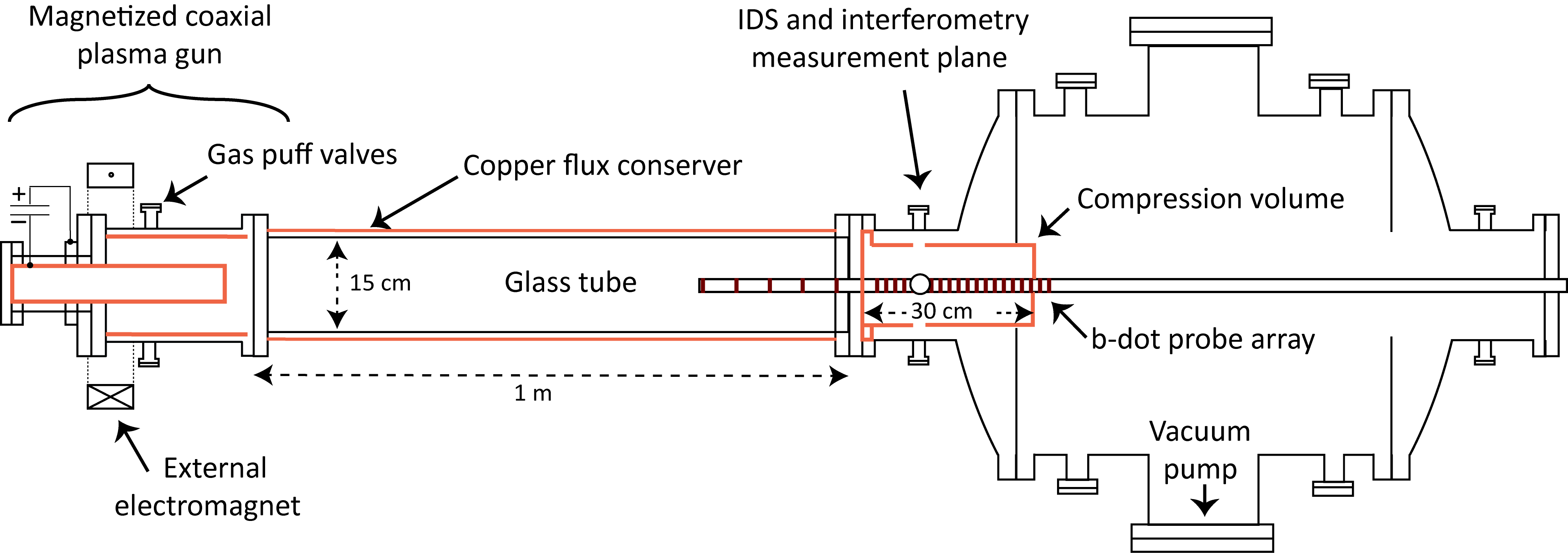}
\end{center}
\caption{Schematic of the SSX device in the wind tunnel configuration. Taylor state plasmas are launched by the magnetized coaxial plasma gun (left).  In the stagnation flux conserver, a long $\dot{B}$ probe array is aligned axially to measure magnetic field structure and time of flight velocity. In addition, co-located ion Doppler spectroscopy and HeNe laser interferometry chords are used for measuring the ion temperature and plasma density, respectively at a distance of $1.24~m$ away from the gun.} 
\label{fig:SSX} 
\end{figure*}

The Swarthmore Spheromak Experiment (SSX) has been used in many different configurations for studying several fundamental phenomena such as magnetic reconnection \cite{Cothran2003} and self-organization \cite{Cothran09,Gray13}, MHD turbulence \cite{BrownPSST}, and magnetothermodynamics \cite{KaurPRE,KaurJPP}. In the present configuration, the SSX device features a $\ell \cong 1.5~m$ long, high vacuum chamber in which we generate $n_e \ge 10^{15}~cm^{-3}, ~T_i \ge 20~eV, ~B \le 0.5~T$ hydrogen plasmas (See Figure \ref{fig:SSX}).  The protons are strongly magnetized ($\rho_i \approx 1~mm$ which is small compared to the dimensions of the machine).  The entire set-up is divided into three main sections: $(i)$ plasma source region, $(ii)$ turbulence region and $(iii)$ the compression region. In the first region, a magnetized coaxial plasma gun is installed which generates fully-ionized, magnetized plasma. 

Plasmas are accelerated to high velocity ($\cong 50~km/s$) by large ${\bf {J} \times B}$ forces in the gun ($10^4~N$) with discharge currents up to $100~kA$, acting on small masses ($100~\mu g$).  Plasmas are injected into a highly evacuated ($10^{-8}$ torr), field-free, cylindrical target volume.  For these studies, the cylinder is highly elongated ($\ell/r \approx 20$), and operated at $120^o~C$.  We find that a hot plasma-facing surface tends to reduce accumulation of cold gas.  The volume is bounded by a highly conducting copper shell ($r = 0.08~m$, thickness $3~mm$). We have used inner plasma-facing surfaces of either glass or tungsten.

The plasma ejected out of the gun is tilt-unstable and turbulently relaxes to a twisted magnetic structure \cite{Taylor74,Taylor86,Cothran09,Gray13}. We use the initial relaxation phase to study MHD turbulence, though this work focuses on the fully-formed plasma object. Parameters match those of earlier studies in which the magnetic structure of the object was confirmed by detailed measurements \cite{Cothran09,Gray13}.
We use an axially-oriented $\dot{B}$ probe array to measure magnetic field structure in this study, and find that the structure is consistent with the previous observations. 
The plasma evolves to an equilibrium that is well described by a non-axisymmetric, force-free state (Taylor state) despite finite plasma pressure ($\beta\approx 20 \%$).

We use vacuum ultraviolet (VUV) spectroscopy for line-averaged measurements of $T_e$ \cite{KaurTe18}.
The VUV spectroscopy is installed $0.05~m$ away from the gun (in the turbulence region) and is line integrated over a diameter.  We find that our electron temperature is about $7~eV$ for most of the discharge.  We can calculate an e-folding lifetime of $\tau = \frac{\mu_0}{\lambda^2 \eta} = 30~\mu s$, where $\eta$ is the Spitzer resistivity and $\lambda$ is the Taylor state eigenvalue from $\nabla \times {\bf B} = \lambda {\bf B}$ \cite{KaurTe18}.

Ion temperatures are measured at the far end of the flux conserver in the compression region ($1.24~m$ away from the gun) using an ion Doppler spectroscopy (IDS) system with a 1~MHz cadence \cite{CothranRSI}. 
We measure emission from $C_{III}$ impurity ions and rely on rapid equilibration of protons with the carbon ions ($\tau_{equil} \le 1~\mu s$). Light from the $C_{III}$ $229.687~nm$ line collected from the plasma along a chord is dispersed to $25^{th}$ order on an echelle grating and is recorded using a 16-channel PMT. The time-resolved proton temperature and line-of-sight average velocity are inferred from the observed thermal broadening and Doppler shift of the emission line, respectively. The line-of-sight is across the flow direction and is located near the end wall of the flux conserver where the plasma stagnates. Our typical peak ion temperatures are $T_i \ge 20~eV$.  
We measure a line-averaged plasma density at the same axial location using a HeNe laser interferometer.  Our typical peak densities are $4 \times 10^{15}~cm^{-3}$.  

\section{\label{sec:Results}Experiment and Results}

In prior studies, we tested three candidate equations of state \cite{KaurPRE,KaurJPP}.  One is appropriate for MHD, the other two are the double-adiabatic, or Chew-Goldberger-Low (CGL) equations of state, appropriate for protons with different $T_{\perp}$ and $T_{\parallel}$ \cite{CGL}:

\begin{equation}
\frac{\partial}{\partial t} \left(\frac{P}{n^{\gamma}} \right)= 0
\end{equation}

\begin{equation}
\frac{\partial}{\partial t} \left(\frac{P_{\perp}}{n B} \right)= 0
\end{equation}

\begin{equation}
\frac{\partial}{\partial t} \left(\frac{P_{\parallel} B^2}{n^3} \right)= 0
\end{equation}

The CGL EOS are related to constancy of adiabatic invariants $\mu = W_{\perp}/B$ and $\mathcal{J} = v_{\parallel} \ell$.  We determined that equation 3 featuring $P_{\parallel}$ best fit our data.  We briefly describe the experimental procedure in what follows.

\begin{figure*}[!h]
\begin{center}
\includegraphics[width=3in]{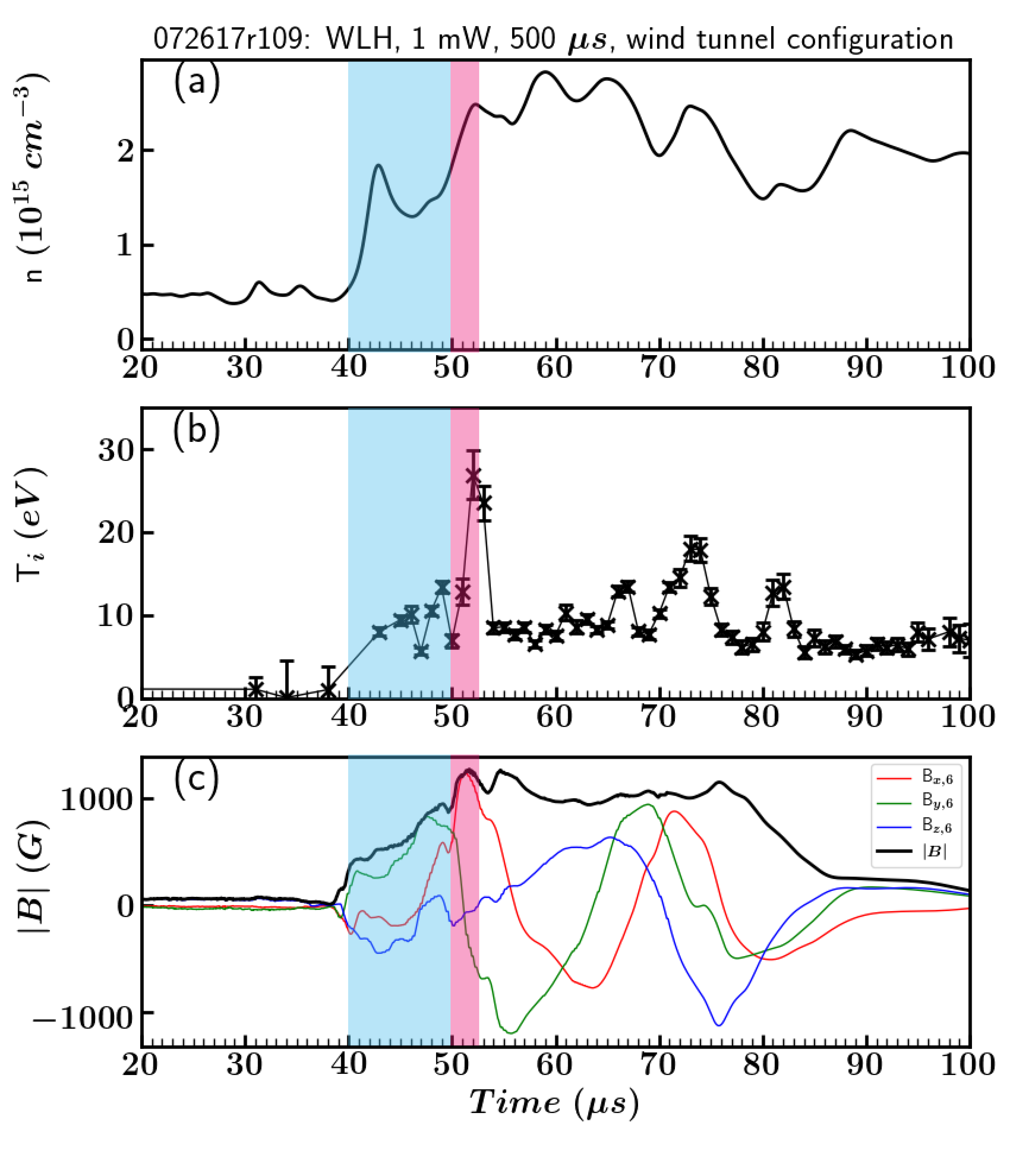}
\end{center}
\caption{Heating event.  Plot of (a) density $n_e$, (b) proton temperature $T_i$, and (c) modulus of the magnetic field $\bf{B}$ measured at the same location in the compression region, highlighting a compressive heating event in pink.  The blue bar indicates the entrance of the Taylor state into the stagnation flux conserver.  For this event, we see a jump in density of about 50\%, and a jump in ion temperature of almost a factor of four.  Combinations of these data as a function of time are used to test EOS models.} 
\label{fig:nBT} 
\end{figure*}

In Figure \ref{fig:nBT} we present a typical time trace from the compression region, highlighting a candidate compression event (pink bar).  During these events, algebraic combinations of these dynamical variables (such as the functional forms above) are used as test model equations of state.  We extract density, temperature, and magnetic field during the compression time.  A viable equation of state would remain constant during the event.

\begin{figure*}[!h]
\begin{center}
\includegraphics[width=3in]{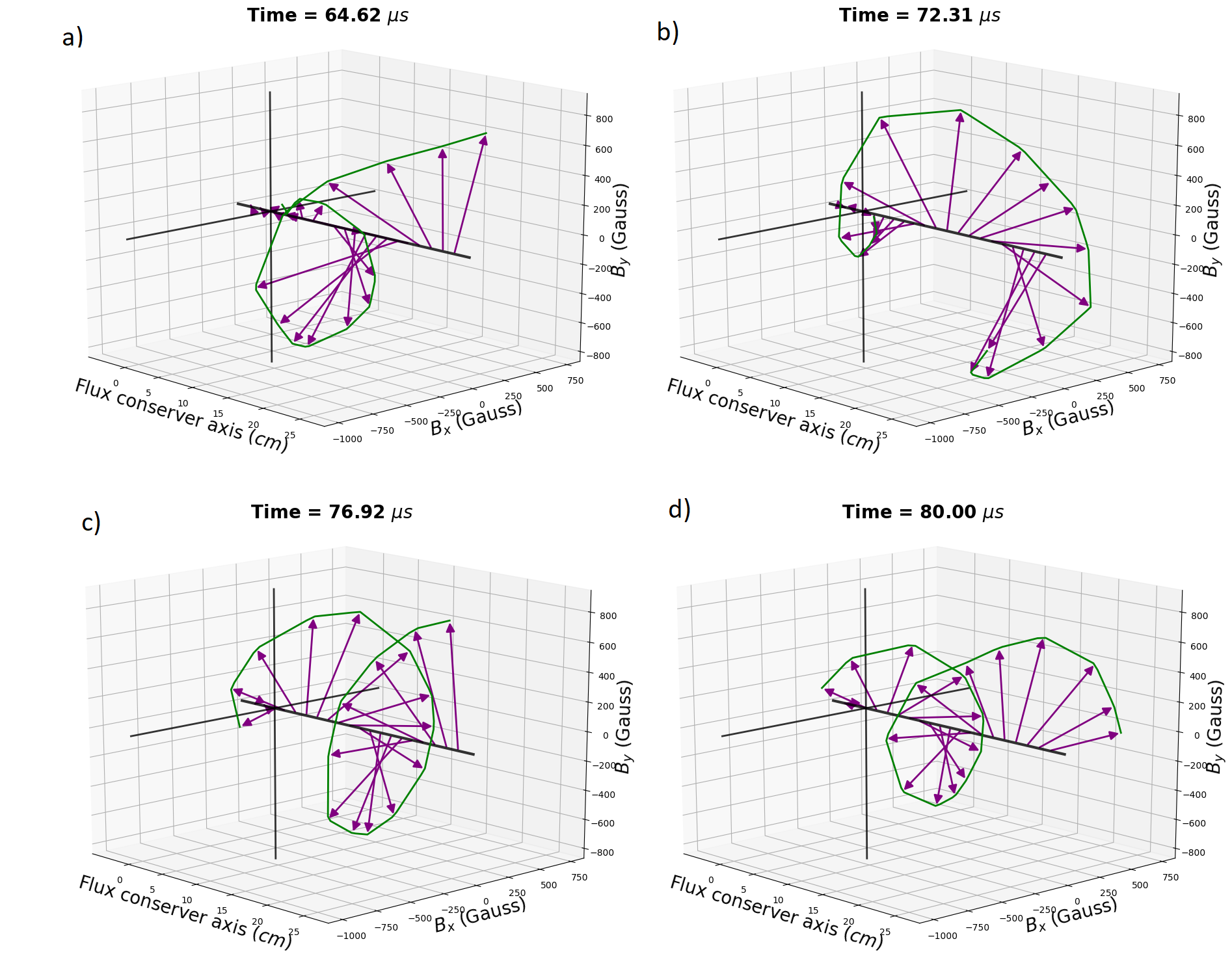}
\end{center}
\caption{Compression.  Plot of the axial structure of the twisted Taylor state as it compresses against the back wall of the stagnation flux conserver.  Note that initially only one lobe is present in the flux conserver ($64~\mu s$), but eventually, 1.5 lobes fill the volume ($80~\mu s$).   It is from this data that we measure the length and volume of the object.} 
\label{fig:twist} 
\end{figure*}

\begin{figure*}[!h]
\begin{center}
\includegraphics[width=3in]{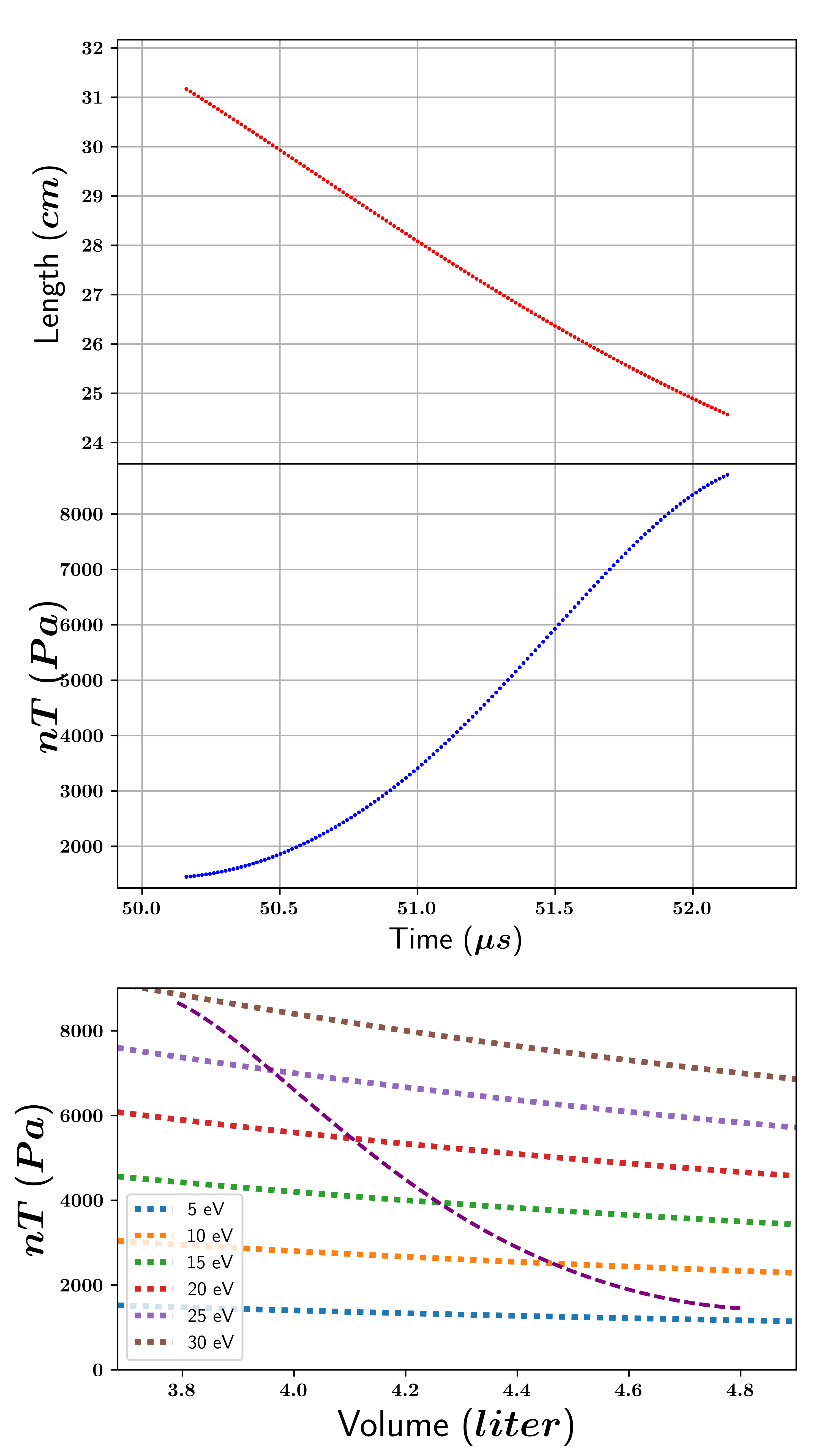}
\end{center}
\caption{PV diagram. We require that the length of the object compresses by at least $10\%$, and that the event persists for at least $1~\mu s$.  In this event, we see a compression of about $20\%$ (from $30$ to $24~cm$).  We also require a transition from a lower isotherm to a higher one.  This event shows a temperature increase of a factor of four.} 
\label{fig:PV} 
\end{figure*}

The axial compression of the Taylor state is depicted in Figure \ref{fig:twist}.  Note that the twisted Taylor state behaves like a spring.  From this data we can extract the axial wavenumber of the structure using wavelet analysis, and therefore obtain the length and volume as a function of time.  We construct a PV diagram using the measured volume (Figure \ref{fig:PV}).

\begin{figure*}[!h]
\begin{center}
\includegraphics[width=3in]{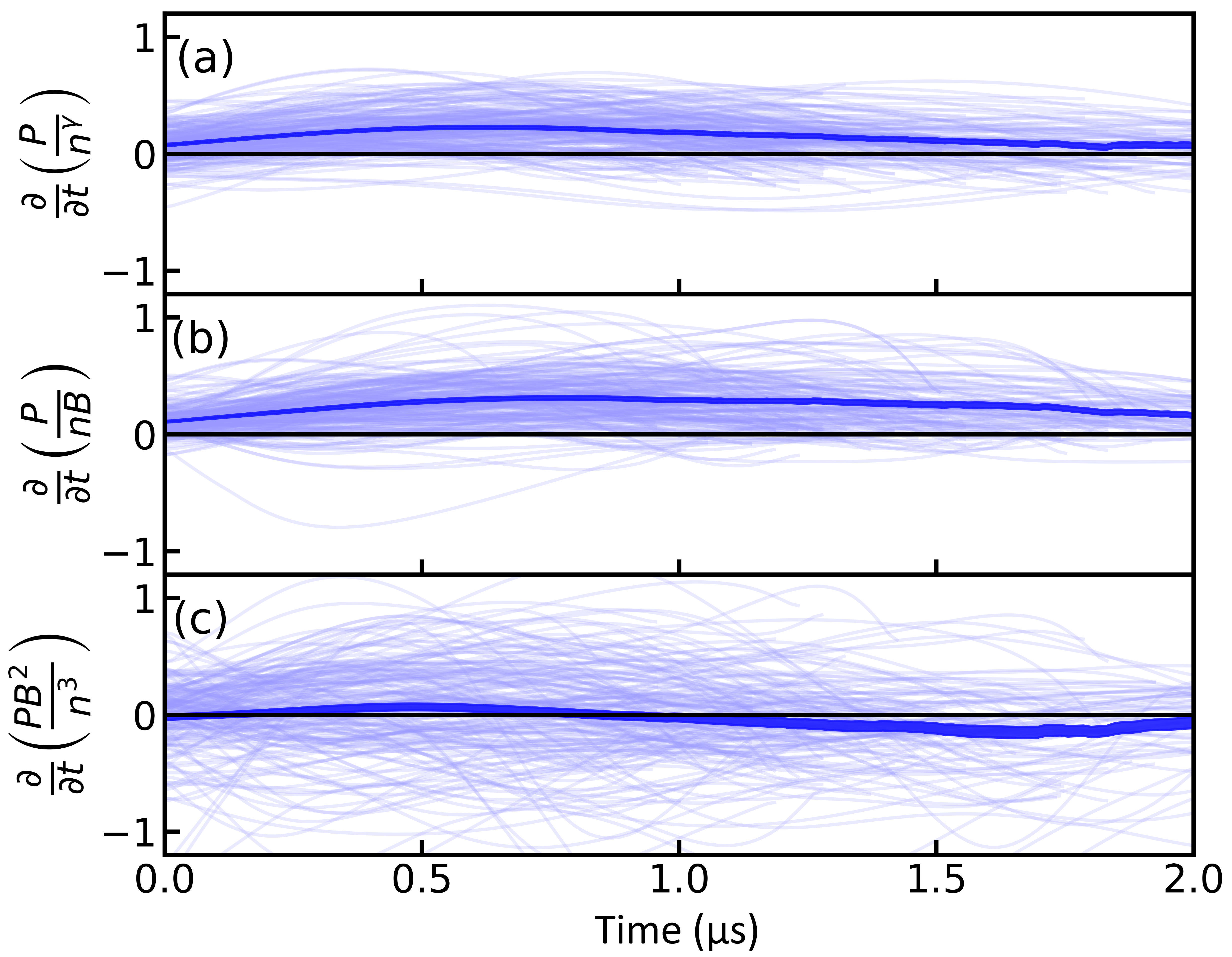}
\end{center}
\caption{Model equations of state.  Nearly 200 distinct compressive heating events are plotted here in light blue, for three candidate EOS models.  The standard error is depicted as a darker blue band.  Since there are 200 events, the standard error is about a factor of $\sqrt{200}$ tighter than the standard deviation. (a) MHD EOS, (b) CGL perpendicular EOS, and (c) CGL parallel EOS.  Note in particular that the MHD EOS is not constant over the heating intervals.  It appears that the parallel CGS EOS is a better model.} 
\label{fig:EOS} 
\end{figure*}

Valid compression events have three criteria. First, the compression is more that 10\%.  Second, the event lasts more than $1~\mu s$ (less than, but on the order of an Alfv\'en time).  Third, there should be ion heating as demonstrated by a transition to a higher isotherm on the PV diagram.  Candidate EOS are displayed in Figure \ref{fig:EOS} for nearly 200 compression events.  Because SSX plasmas relax to an equilibrium described by MHD (eg \cite{Cothran09,Gray13}), one might expect the thermodynamics to be described by the corresponding EOS.  Instead, we find that the parallel CGL EOS best fits our data.

Our measurement of $T_i$ is insensitive to the direction of magnetic field.  The IDS diagnostic is line-averaged over a diametrical chord where the direction of the magnetic field changes significantly, so we mix both $T_{\perp}$ and $T_{\parallel}$ information along a chord.  In addition, the IDS collection chord is several $mm$ in diameter \cite{CothranRSI} so even when the magnetic field is fortuitously aligned with the optics, the sampling volume is over several proton Larmor orbits.  We are considering other techniques to resolve $T_{\perp}$ and $T_{\parallel}$, but the results presented here are averaged over both components.  In light of this averaging, it is particularly interesting that the parallel CGL EOS best fits our data.

\section{\label{Engine}A Fusion Engine}

Using our measured equation of state, we can hypothesize how a Taylor state would perform as a working fluid in a fusion engine.  We imagine that we are able to compress the Taylor state by a factor of 10 or more, and that the magnetofluid obeys our candidate equation of state: $$\frac{\partial}{\partial t} \left(\frac{P_{\parallel} B^2}{n^3} \right)= 0.$$  We make three assumptions.  First, that the compression occurs faster than the magnetic decay time: $\tau_{compress} \ll \tau_B$.  Second, we assume that the total particle number is conserved during compression, so the density increases inversely with volume: $n \propto 1/V$.  Finally, that magnetic energy is conserved.  This means that $W_{mag} = \int (B^2/2 \mu_0) dV$ is a constant, so $B^2$ also increases inversely with volume: $B^2 \propto 1/V$.

\begin{figure*}[!h]
\begin{center}
\includegraphics[width=3in]{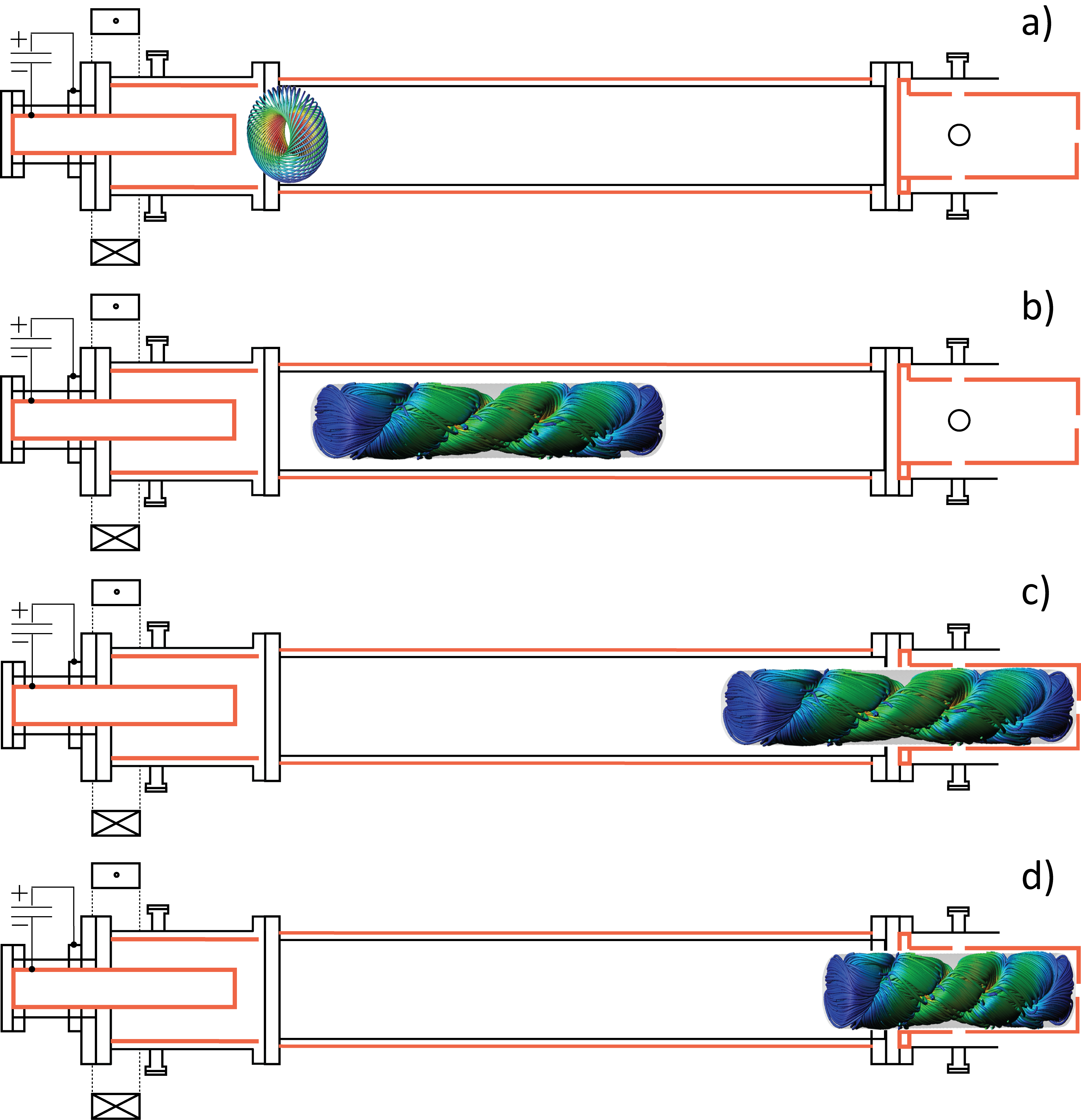}
\end{center}
\caption{Fusion engine.  Axial compression of a Taylor state, in this case, depicting the $30\%$ compression we typically see at SSX.  A fusion engine would need at least a factor of 10 compression.  Radial compression is also a possibility, but would require an imploding liner or collapsing liquid metal wall.} 
\label{fig:engine} 
\end{figure*}

In Figure \ref{fig:engine}, we show a cartoon of axial compression.  We imagine a convergence ratio of 10 or more could be performed either axially or radially.  Radial compression could be done either with a collapsing liquid metal wall \cite{Laberge09} or an imploding liner \cite{Gomez14,Gomez15}.  Axial compression could be done as we show here at SSX, but the Taylor state would have to be accelerated to super-Alfv\'enic speeds.

From this analysis, we conclude:
$$\frac{P_{\parallel} B^2}{n^3} = \frac{n k T_{\parallel} B^2}{n^3} \propto \frac{T_{\parallel} B^2}{n^2} = \rm{constant}$$  Since $B^2$ and $n$ both scale like $1/V$, we find that $T_{\parallel} V = \rm{constant}$, so that a decrease in volume by a factor of 10, results in an increase in temperature by a factor of 10.  Note that this is much more favorable scaling than is predicted by the MHD equation of state: 
$$\frac{P}{n^{\gamma}} = \rm{constant}$$ for which the prediction is that $T V^{\gamma-1} = T V^{2/3} = \rm{constant}.$  A convergence ratio of 10 would bring fusion fuel from $1~keV$ to $10~keV$ while increasing the density by a factor of 10.

\section{\label{Sum}Summary}

We have presented results from SSX experiments on the thermodynamics of compressed magnetized plasmas.  The key aspects of the experiment are to compress rapidly so the physics is adiabatic, yet quasi-statically so that no shocks develop (practically speaking, this means $\tau_{compress} \le \tau_{Alf}$).  We identify a particular equation of state, associated with the CGL parallel adiabatic invariant.

\section{\label{sec:Ack}Acknowledgements} This work was supported by the DOE Advanced Projects Research Agency (ARPA) ALPHA program project DE-AR0000564.  The authors wish to acknowledge the support and encouragement of ARPA program managers Dr. Patrick McGrath and Dr. Scott Hsu.  We would like to particularly acknowledge undergraduate student contributions from Katie Gelber, Nick Anderson, Hari Srinivasulu, Emma Suen-Lewis, Luke Barbano, and Jaron Shrock, and technical discussions with colleagues David Schaffner, Adam Light, and Simon Woodruff.  Technical support from Steve Palmer and Paul Jacobs at Swarthmore for SSX is also gratefully acknowledged.

\begin{thebibliography}{99}

\bibitem{BrownJPP}
M. R. Brown and D. A. Schaffner, ``SSX MHD plasma wind tunnel'', {\em J. Plasma Physics} {\bf 81}, 345810302 (2015).

\bibitem{BrownPSST}
M. R. Brown and D. A. Schaffner, ``Laboratory sources of turbulent plasma: a unique MHD plasma wind tunnel'', {\em Plasma Sources Science and Technology}, invited review, {\bf 23}, 063001 (2014).

\bibitem{Taylor74}
J. B. Taylor, ``Relaxation of Toroidal Plasma and Generation of Reverse Magnetic Fields'', {\em Phys. Rev. Lett.} {\bf 33}, 1139 (1974).

\bibitem{Taylor86}
J. B. Taylor, ``Relaxation and magnetic reconnection in plasmas'', {\em Rev. Mod. Phys.} {\bf 58}, 741 (1986).

\bibitem{Cothran09}
C. D. Cothran, M. R. Brown, T. Gray, M. J. Schaffer, and G. Marklin, ``Observation of a Helical Self-Organized State in a Compact Toroidal Plasma'', {\em Phys. Rev. Letters} {\bf 103}, 215002 (2009).

\bibitem{Gray13}
T. Gray, M. R. Brown, and D. Dandurand, ``Observation of a Relaxed Plasma State in a Quasi-Infinite Cylinder'', {\em Phys. Rev. Letters} {\bf 110}, 085002 (2013).

\bibitem{KaurPRE}
Kaur, M., Barbano, L. J., Suen-Lewis, E. M., Shrock, J. E., Light, A. D., Brown, M. R., and Schaffner, D. A., ``Measuring the equations of state in a relaxed magnetohydrodynamic plasma", {\em Phys. Rev. E} {\bf 97}, 011202 (2018).

\bibitem{KaurJPP}
Kaur, M., Barbano, L., Suen-Lewis, E., Shrock, J., Light, A., Schaffner, D.A., Brown, M.R., Woodruff, S., Meyer, T., ``Magnetothermodynamics: Measurements of the thermodynamic properties in a relaxed magnetohydrodynamic plasma", {\em Journal of Plasma Physics} {\bf 84}, 905840114 (2018).

\bibitem{Lindemuth95}
I. R. Lindemuth, et al. ``Target Plasma Formation for Magnetic Compression/Magnetized Target Fusion'', {\em Phys. Rev. Lett.} {\bf 75}, 1953 (1995).

\bibitem{Wurden16}
G. A. Wurden, et al., ``Magneto-Inertial Fusion'', {\em Journal of Fusion Energy} {\bf 35}, 69 (2016).

\bibitem{Degnan13}
J. H. Degnan, et al., ``Recent magneto-inertial fusion experiments on the field reversed configuration heating experiment'', {\em N. Fusion} {\bf 53}, 093003, (2013).

\bibitem{Laberge09}
M. Laberge, ``Experimental Results for an Acoustic Driver for MTF'', {\em Journal of Fusion Energy} {\bf 28}, 179 (2009).

\bibitem{Gomez14}
M. R. Gomez, et al., ``Experimental Demonstration of Fusion-Relevant Conditions in Magnetized Liner Inertial Fusion'', {\em Phys. Rev. Lett.} {\bf113}, 155003 (2014).

\bibitem{Gomez15}
M. R. Gomez, et al., ``Demonstration of thermonuclear conditions in magnetized liner inertial fusion experiments'', {\em Physics of Plasmas} {\bf22}, 056306 (2015).

\bibitem{Cothran2003}
Cothran, C. D., M. Landreman, M. R. Brown, and W. H. Matthaeus (2003), ``Three-dimensional structure of magnetic reconnection in a laboratory plasma", {\em Geophys. Res. Lett.} {\bf 30}, 1213 (2003).

\bibitem{KaurTe18}
Kaur, M., K. D. Gelber, A. D. Light, and M. R. Brown, ``Temperature and Lifetime Measurements in the SSX Wind Tunnel'', {\em Plasma} {\bf 1(2)}, 229-241 (2018).

\bibitem{CothranRSI}
C. D. Cothran, J. Fung, M. R. Brown, and M. J. Schaffer, ``Fast, High Resolution Echelle Spectroscopy of a Laboratory Plasma'' {\em Review of Scientific Instruments} {\bf 77}, 063504 (2006).

\bibitem{CGL}
G. F. Chew, M. L. Goldberger, and F. E. Low, ``The Boltzmann equation and the one-fluid hydromagnetic equations in the absence of particle collisions'', {\em Proc. R. Soc.
Lond. Ser. A} {\bf 236}, 112 (1956).

\end{thebibliography}
\end{document}